\documentclass[submission, Proceedings]{SciPost}

\hypersetup{
    colorlinks,
    linkcolor={red!50!black},
    citecolor={blue!50!black},
    urlcolor={blue!80!black}
}

\usepackage[bitstream-charter]{mathdesign}
\DeclareSymbolFont{usualmathcal}{OMS}{cmsy}{m}{n}
\DeclareSymbolFontAlphabet{\mathcal}{usualmathcal}

\newcommand{\ttbar}{$\mathrm{t\bar{t}}$}
\newcommand{\ttgamma}{$\mathrm{t\bar{t}\gamma}$}
\newcommand{\ttZ}{$\mathrm{t\bar{t}Z}$}
\newcommand{\fbinv}{$\mathrm{fb^{-1}}$}
\newcommand{\yukawa}{$\mathrm{Y_{t}}$}
\newcommand{\mttbar}{$\mathrm{m_{t\bar{t}}}$}

\begin{document}

\begin{center}{\Large \textbf{
Recent results on top quark mass and properties and rare/anomalous top quarks interactions in CMS\\
}}\end{center}

\begin{center}
S. Wuchterl\textsuperscript{$\star$}
on behalf of the CMS Collaboration
\end{center}

\begin{center}
{\bf} Deutsches Elektronen-Synchrotron (DESY), Hamburg, Germany
\\
* sebastian.wuchterl@cern.ch
\end{center}

\begin{center}
\today
\end{center}


\definecolor{palegray}{gray}{0.95}
\begin{center}
\colorbox{palegray}{
  \begin{tabular}{rr}
  \begin{minipage}{0.1\textwidth}
    \includegraphics[width=22mm]{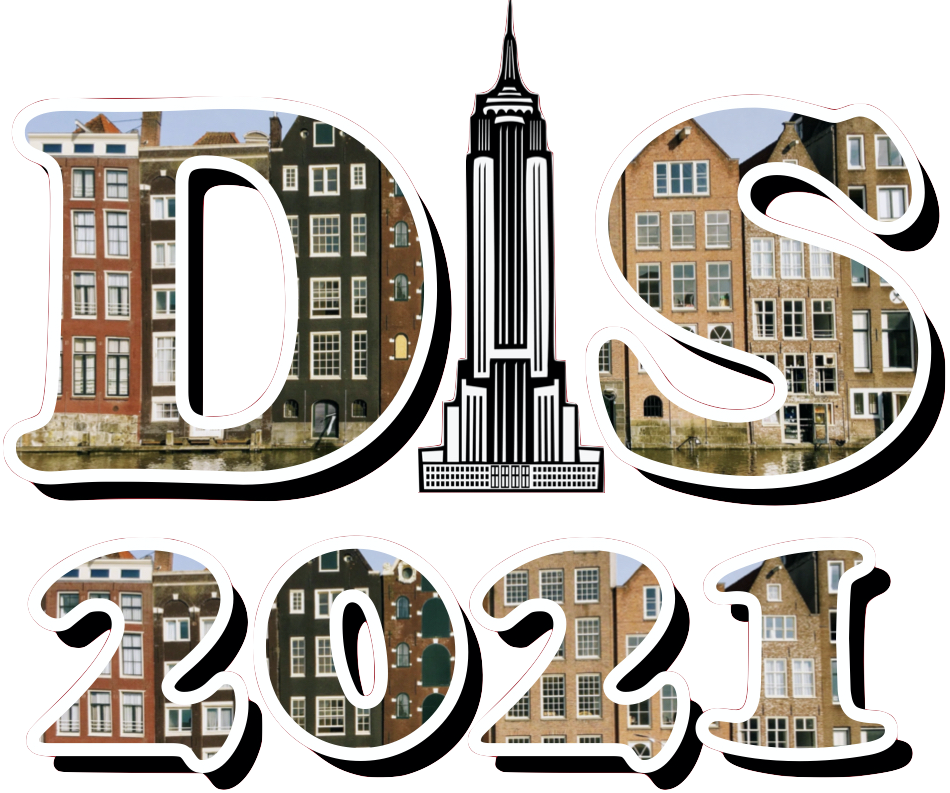}
  \end{minipage}
  &
  \begin{minipage}{0.75\textwidth}
    \begin{center}
    {\it Proceedings for the XXVIII International Workshop\\ on Deep-Inelastic Scattering and
Related Subjects,}\\
    {\it Stony Brook University, New York, USA, 12-16 April 2021} \\
    \doi{10.21468/SciPostPhysProc.?}\\
    \end{center}
  \end{minipage}
\end{tabular}
}
\end{center}

\section*{Abstract}
{\bf
The amount of data collected by the CMS experiment at the CERN LHC in the second data taking period provides the possibility to study absolute and differential cross sections of top quark interaction processes with high precision. Utilizing these results, fundamental standard model parameters such as the top quark mass are extracted. Moreover, given standard model measurements are interpreted in the context of Beyond the standard model theories.
Recent results on top quark properties and rare or anomalous top quark interactions by CMS are presented in these proceedings. Furthermore, the implications of experimental results to effective field theory are discussed. 
}


\section{Introduction}
\label{sec:intro}
The standard model of particle physics (SM) describes successfully a variety of processes in a wide energy range up to several~TeV. Its properties and parameters have been investigated and determined by different experiments such as the CMS experiment~\cite{bib:CMS} at the CERN LHC over the past decades. However, open questions such as the origin of dark matter cannot be explained by the SM, and it is indicated that the SM could be the low energy approximation of a more general higher order theory.
The mass of the most massive elementary particle, the top quark, suggests that the top quark takes a special role within the SM. Its Yukawa coupling to the Higgs boson is close to unity and hence it implies its particular importance in the electroweak symmetry breaking. Therefore, the precise study of processes involving top quarks can shed a light both on beyond the SM (BSM) physics but also solve the question on the stability of the electroweak vacuum by performing precise measurements of the top quark mass.

At hadron colliders such as the CERN LHC, top quarks are produced with a high production rate. They are produced predominantly in pairs (\ttbar\ production), but also in single top production or associated with light or heavy bosons (e.g. \ttZ\ and \ttgamma). Thus, with the extensive amount of data collected in the second data taking period by the CMS experiment, the introduced processes can be investigated with a precision that is compatible or more precise than recent theoretical calculations. Because the top quark decays before performing bound states, it provides also a unique way to study quark properties or to extract other SM parameters.

\section{Top quark mass measurements}
\label{sec:topmass}
The value of the top quark mass, $m_{t}$, can be measured in mainly two different ways, which involve a different theoretical understanding and can be classified as direct and indirect measurements. Whereas direct measurements have reached a precision up to $0.5\,\mathrm{GeV}$, they lack a well defined theoretical definition. They are based on the kinematic reconstruction of the top quark decay products and thus rely on inherent features and the probabilistic modeling of multi purpose Monte Carlo (MC) generators. On the other hand, indirect measurements extract the top quark mass in well defined renormalization schemes, such as the pole or modified minimal subtraction ($\mathrm{\overline{MS}}$) renormalization scheme, by measuring observables with a direct sensitivity to the top quark mass. The relation between the top quark MC mass, $\mathrm{m_{t}^{MC}}$, from direct measurements, is usually related to the pole mass with an additional uncertainty of the order of $1\,\mathrm{GeV}$ motivated by the uncertainty associated to the modeling of MC generators~\cite{bib:massTheoHoang,bib:massTheoNason}.

\begin{figure}[h]
\centering
\includegraphics[width=0.48\textwidth]{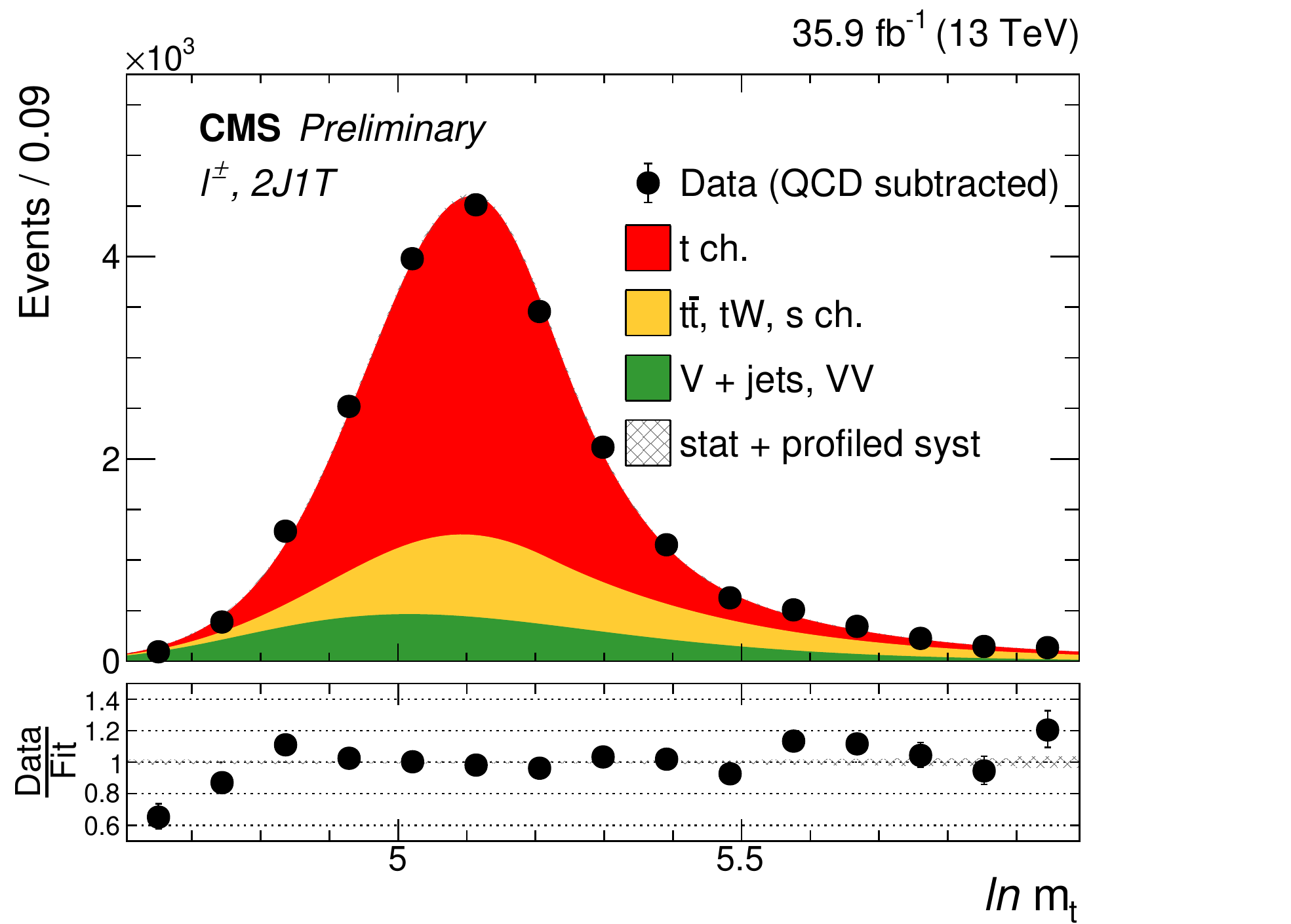}
\includegraphics[width=0.50\textwidth]{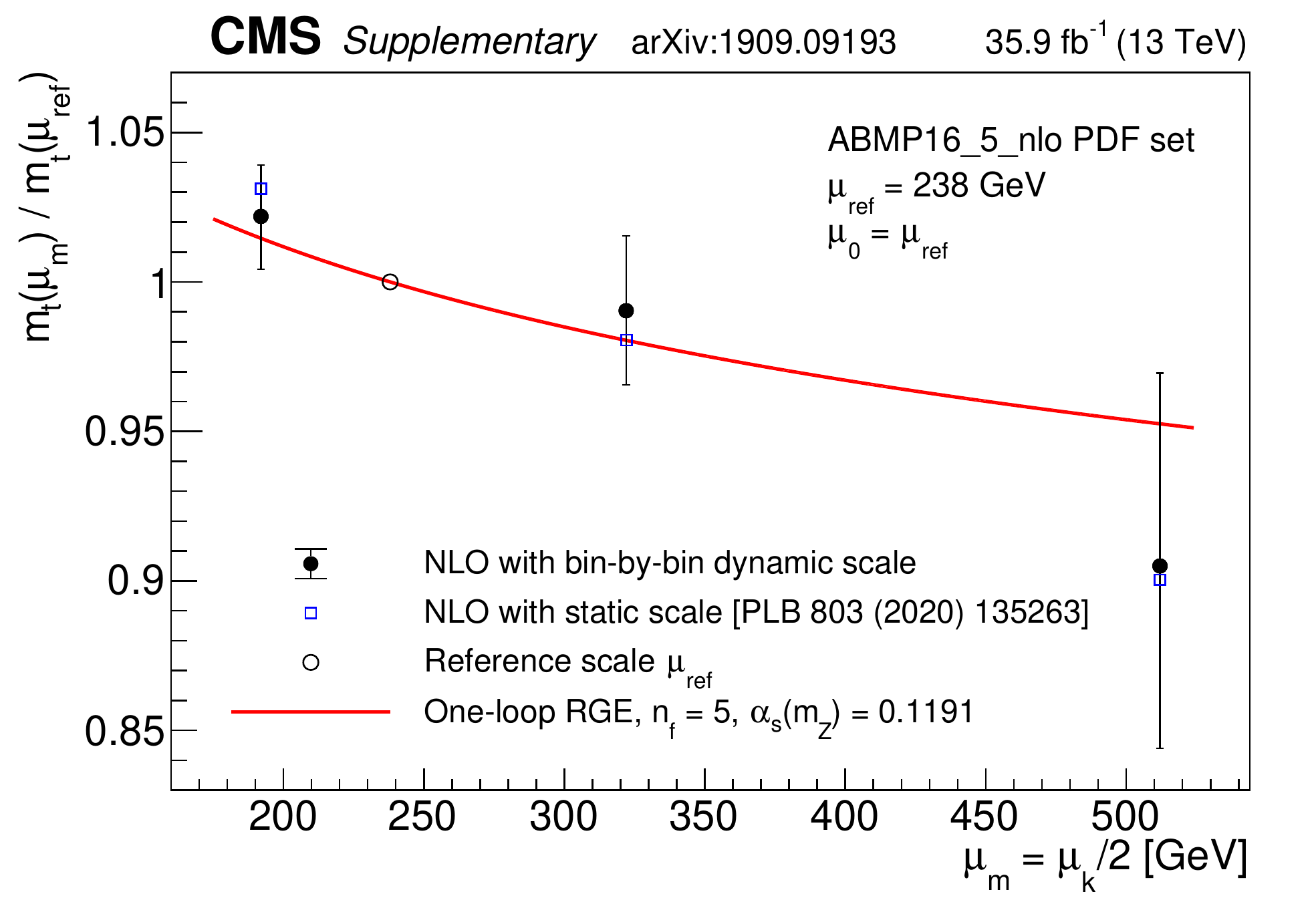}
\caption{Reconstructed data as a function of the natural logarithm of the reconstructed top quark mass together with a functional fit (left)~\cite{bib:singletopPAS}. Measured scale dependence of the top quark mass as a function of the scale $\mathrm{\mu_{m}}=\mathrm{m_{t\bar{t}}}/2$ (right)~\cite{bib:runningPlot,bib:running}.}
\label{fig:mass1}
\end{figure}

The top quark mass was measured by the CMS collaboration recently in the single top t-channel using $35.9$~\fbinv\ of pp collision data~\cite{bib:singletopPAS}. By separating events based on the lepton charge in the analyzed leptonic decay channel of the top quark, both the mass of the top quark and the top antiquark are additionally measured individually by a functional fit, as it is shown in Figure~\ref{fig:mass1} left for the combined fit. By making use of multivariate analysis techniques, the signal sensitivity is enhanced and the combined top quark/antiquark mass is measured to be $172.13 \pm 0.7\, \mathrm{GeV}$. The extracted ratio between the quark and antiquark is $0.995\pm0.006$, and their difference is found to be $0.83_{-1.01}^{+0.77}$. The uncertainties are dominated mainly by the jet energy scale, final state radiation, and color reconnection modeling uncertainties. This is the first time that $\mathrm{m_{t}^{MC}}$ was measured with a precision below $1\,\mathrm{GeV}$ in the single top phase space.

By measuring the top quark mass in the boosted regime of decaying top quarks in the \ttbar\ pair production process~\cite{bib:boosted}, the relation between direct and indirect measurements can be probed~\cite{bib:boostedHoang}. The top quark mass is extracted to be $172.6 \pm 1.9 \mathrm{GeV}$ by comparing the unfolded differential cross section as a function of the jet mass at particle level to theoretical predictions. The same amount of pp collision data as in the aforementioned analysis was used. The reported results significantly improved with respect to former measurements by the usage of the Xcone jet clustering algorithm. It was used for the first time at LHC to improve the resolution of the jet mass at particle and detector levels.

The first experimental investigation of the scale dependence (running) of the top quark mass~\cite{bib:running} is performed by analyzing \ttbar\ events using also $35.9$~\fbinv\ of pp collision data. The top quark mass running in the $\mathrm{\overline{MS}}$ scheme is a fundamental effect of quantum chromodynamics and analogous to the running of the strong coupling constant. Through a profiled maximum likelihood fit to final state observables, the differential cross section as a function of the invariant mass of the \ttbar\ system (\mttbar) is measured at parton level. From a comparison of the measured cross section to fixed order theory prediction, the running is probed up to a scale of $1\,\mathrm{TeV}$ and found to agree with the SM within $1.1\mathrm{\sigma}$.

\section{Measurements of top quark properties}
\label{sec:properties}
Utilizing the special properties of the top quark, fundamental SM parameters can be extracted by investigating processes involving top quarks. In the presented measurement, the top quark Yukawa coupling \yukawa\ is determined by a likelihood fit to final state distributions of \ttbar\ events using $137$\fbinv\ of pp collision data~\cite{bib:yukawa}. Higher order electroweak effects affecting the Yukawa coupling are modeled using HATHOR~\cite{bib:hathor} predictions, and proxy variables are fitted to mitigate the dependence of the missing transverse momentum resolution in the leptonic decay channel. The best fit value for \yukawa\ yields $\mathrm{Y_{t}}=1.16^{+0.24}_{-0.35}$ and an upper limit of $Y_{t}<1.5$ is extracted at 95\%~CL.

By separating single top t-channel signal events into multiple contributions depending on the bottom-bottom and bottom-light quark interactions at both $\mathrm{tWb}$ vertices, the CKM matrix elements $\mathrm{V_{tb}}$, $\mathrm{V_{Vtd}}$, and $\mathrm{V_{ts}}$ are measured the first time simultaneously and model independent~\cite{bib:CKM}. In pp collision events corresponding to an integrated luminosity of $35.9$~\fbinv, multivariate discriminator response observables are fitted in a likelihood fit to determine the signal strengths of the individual single top t-channel modes. A three-fold interpretation of the measured signal strength parameters is provided, setting limits on the CKM matrix elements under SM unitary assumption ($\mathrm{|V_{tb}|>0.970}$, $\mathrm{|V_{td}|^2 + |V_{ts}|^2 < 0.057}$), allowing for more quark families ($\mathrm{|V_{tb}|=0.988\pm0.051}$, $\mathrm{|V_{td}|^2 + |V_{ts}|^2 = 0.06\pm0.06}$), or leaving the top quark width unconstrained ($\mathrm{|V_{tb}|=0.988\pm0.024}$, $\mathrm{|V_{td}|^2 + |V_{ts}|^2 < 0.06\pm0.06}$, $\mathrm{\frac{\Gamma_{t}^{obs}}{\Gamma_{t}}=0.99\pm0.42}$). 

\begin{figure}[h]
\centering
\includegraphics[width=0.54\textwidth]{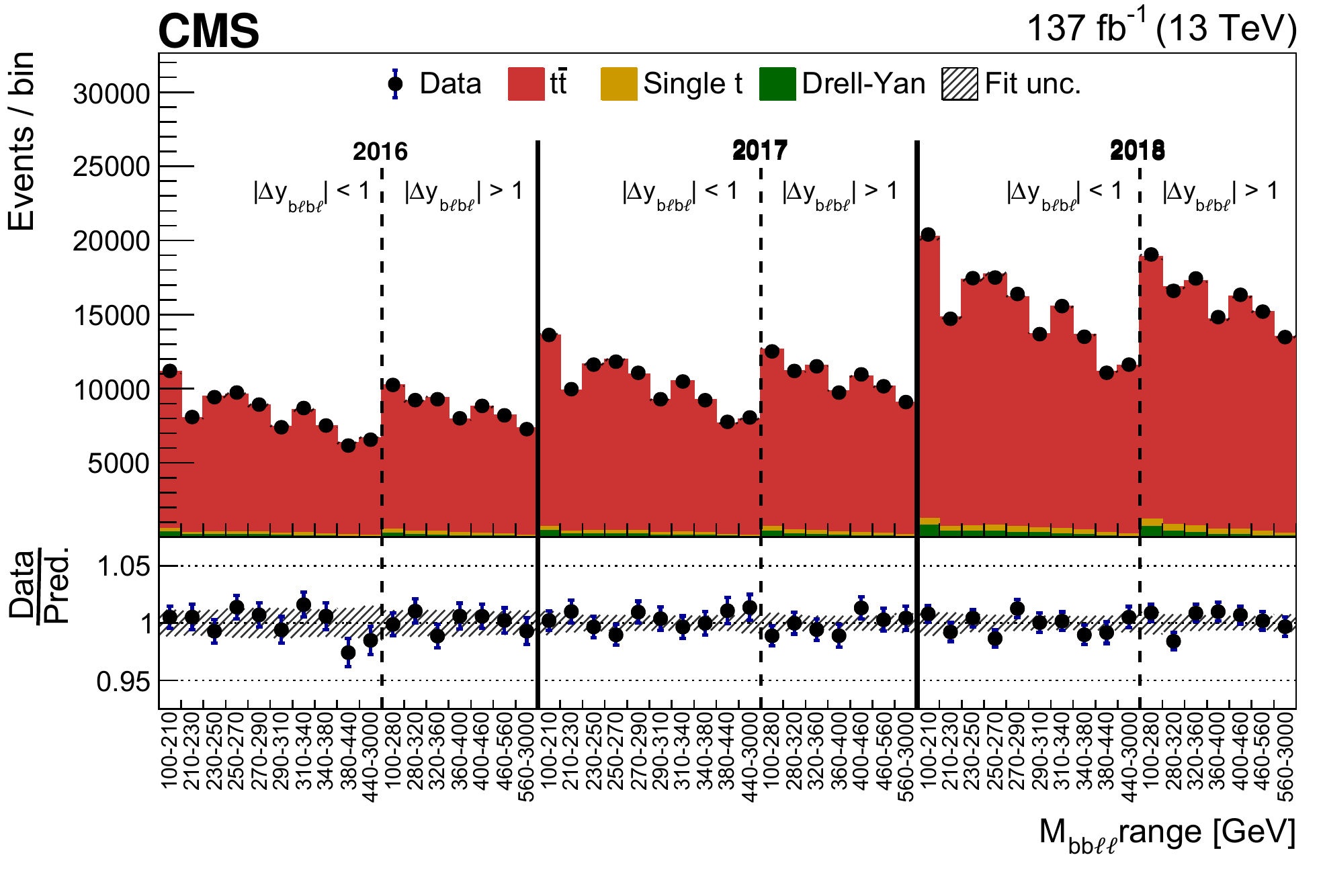}
\includegraphics[width=0.45\textwidth]{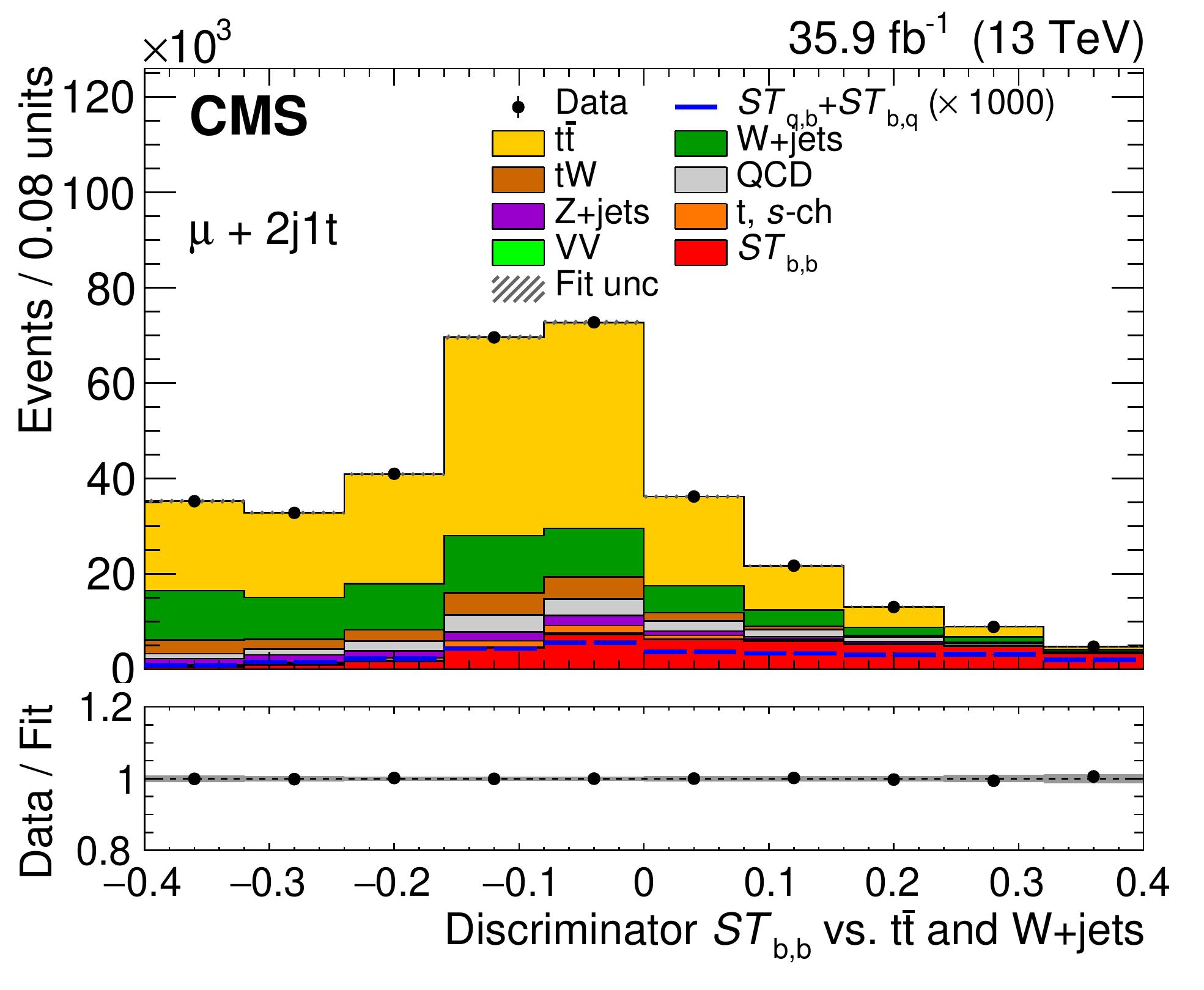}
\caption{Comparison between measured data and SM expectation in different bins of mass and rapidity bins for a post-fit result of $\mathrm{Y_{t}}=1.16$ (left)~\cite{bib:cp}. Distribution of the multivariate discriminator comparing data to simulation after the fit procedure (right)~\cite{bib:CKM}.}
\label{fig:prop1}
\end{figure}

\begin{figure}[h]
\centering
\includegraphics[width=0.99\textwidth]{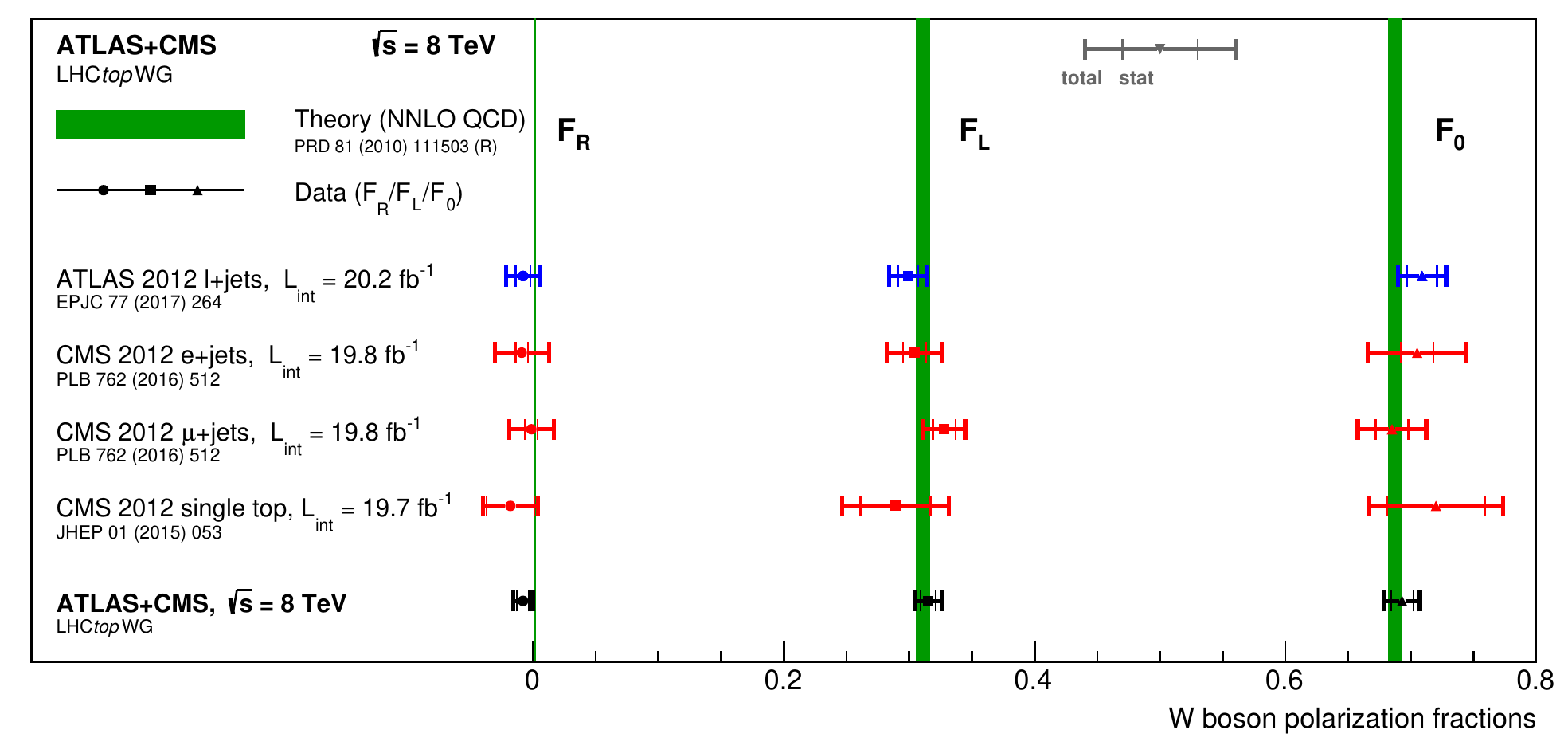}
\caption{Summary plot of the individual measurements of the W polarization fractions and their combination~\cite{bib:W}.}
\label{fig:prop2}
\end{figure}

Alike, the polarization of the W boson can be extracted from differential cross sections of \ttbar\ and t-channel single top production~\cite{bib:W}. Several analyses have been published by the ATLAS and CMS collaborations for the analyzed pp collision data at 8 TeV, and here a combination of those results is reported. Correlations of systematic uncertainties have been studied between the analyses, and by using the best linear unbiased estimator method, combined results for the right handed, left handed, and longitudinal W boson polarization fractions $\mathrm{F_{R}}$, $\mathrm{F_{L}}$, and $\mathrm{F_{0}}$ are determined. As can be seen in Figure~\ref{fig:prop2}, with an overall relative improvement of $25-30\%$, the combination yields much more precise results than the individual measurements alone.

Because top quark pairs are produced mainly via gluon fusion at the CERN LHC, measuring the forward-backward asymmetry ($\mathrm{A_{FB}}$) in \ttbar\ production is a considerable larger challenge compared to proton-antiproton colliders as the TEVATRON. In the results presented~\cite{bib:forward}, pp collision data corresponding to $35.9$ \fbinv\ are analyzed in the boosted and resolved decay topologies. By means of a likelihood fit to three final state observables, \mttbar, the scattering angle $\mathrm{c^{*}}$, and the longitudinal momentum fraction of the top quarks $\mathrm{x_{F}}$, gluon and quark initiated production modes are separated and the asymmetry is extracted. The resulting value for $\mathrm{A_{FB}}$ of $0.048^{+0.021}_{-0.014}$ is found to be consistent with the SM and an upper limits on the chromomagnetic and chromoeletric dipole moment (CEDM) of the top quark are set.

In a similar approach analyzing the same set of pp collision events, the conservation of CP symmetry in \ttbar\ decays is examined~\cite{bib:cp}. Theoretically, a large fraction of CP violation would be predicted from the CEDM of the top quark. The asymmetry of the Levi-Cita tensors $\mathrm{O_{1}}$ and $\mathrm{O_{3}}$ of the four momenta of the top quarks, charged leptons, and b quarks are directly sensitive to this effect. From their measurement, values for the CEDM ($0.058\pm0.098$ and $-0.01\pm0.092$) of the top quark are determined and are found to be consistent with the SM expectation.

\section{Rare processes \& anomalous couplings}
\label{sec:eft}
One approach to interpret SM measurements is the framework of effective field theory (EFT). By probing anomalous couplings of higher order operators and their dimensionless coupling strengths represented by Wilson coefficients (WCs), a model independent picture of BSM effects can be obtained.
Two measurements performed by CMS investigated the absolute and differential cross section of \ttbar\ associated production with photons or Z bosons, respectively. The data sets analyzed correspond to $77.5$~\fbinv\ for the \ttgamma~\cite{bib:ttgamma} and $137$~\fbinv\ for the \ttZ\ measurement~\cite{bib:ttZ}. Of particular sensitivity to EFT operators affecting \ttgamma\ and \ttZ\ production are the momentum distribution of the Z boson and the photon, and further the angle of the negatively charged lepton in the boson rest frame with respect to the Z boson. Both analyses together set the tightest constraints on the scrutinized couplings to date.

A novel approach was employed in the presented CMS analysis by parametrizing EFT effects from multiple operators simultaneously in detector level observables~\cite{bib:eft}. Multilepton final state events were analyzed using $137$~\fbinv\ pp collision data and the total yields of five associated production modes of the top quark ($\mathrm{t\bar{t}Z}$, $\mathrm{t\bar{t}W}$, $\mathrm{t\bar{t}H}$, $\mathrm{tZq}$, $\mathrm{tHq}$) are fitted. Using signal categories enriched and particularly sensitive to one of the five signals studied, the effect of 16 WCs modifying the signal strengths is probed. By profiling them one by one or all WCs simultaneously, confidence intervals corresponding to two standard deviations are computed. This analysis represents the first important step to a global EFT interpretation in all relevant processes.

\begin{figure}[h]
\centering
\includegraphics[width=0.99\textwidth]{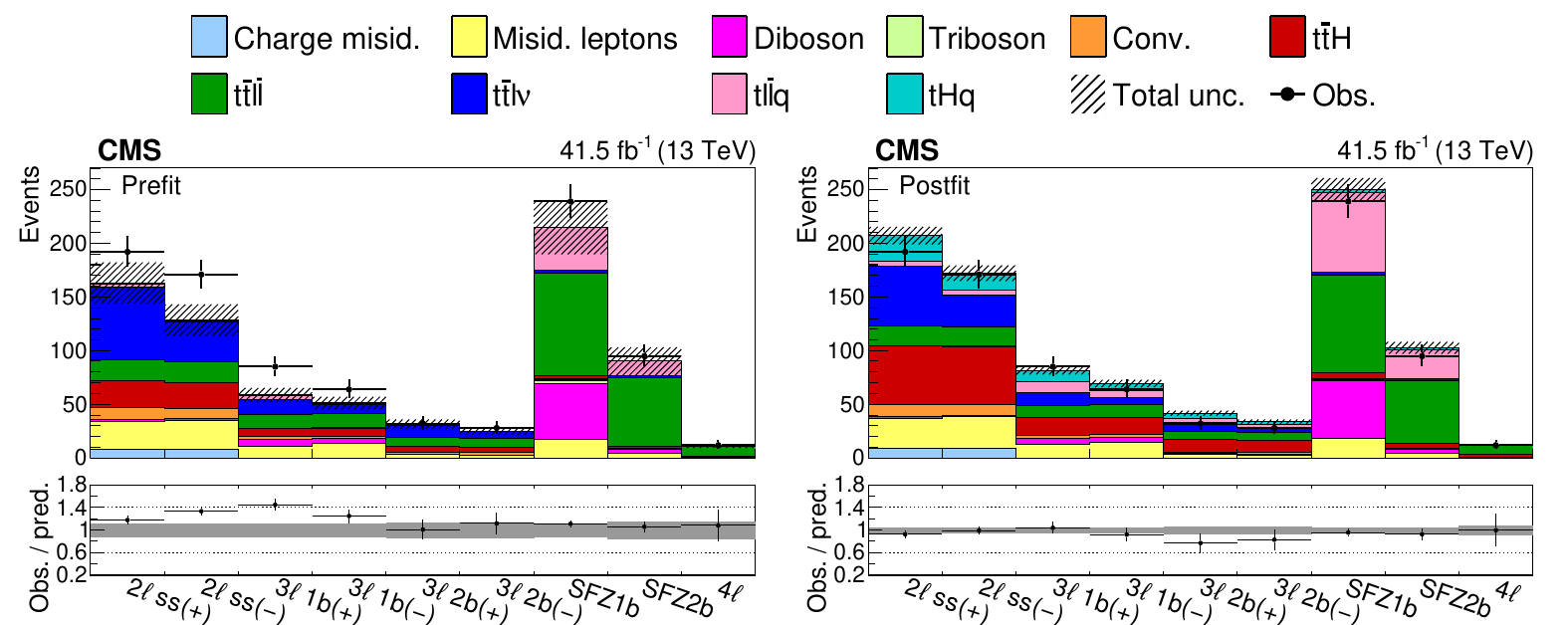}
\caption{Expected signal category yields before (left) and after the fit procedure (right), together with the comparison to measured data~\cite{bib:eft}.}
\label{fig:eft1}
\end{figure}

\section{Conclusion}
Recent results on top quark mass and properties measurements by CMS have been reported. Furthermore, measurements of rare and anomalous top quark interactions have been presented with their implications for physics beyond the standard model.
Significant improvements in direct and indirect measurements of the top quark mass have been achieved, analyzing single-top and top pair production events. Several standard model parameters have been extracted from studying top quark interactions, such as the top quark Yukawa coupling, CKM matrix elements, or W boson polarization fractions. While good agreement with the standard model is observed, interpretations of differential cross section measurements in the effective field theory framework have been highlighted. Together with recent analyses employing novel parametrization techniques, they set the most stringent limits on various anomalous couplings to date.










\bibliography{bibliography.bib}

\nolinenumbers

\end{document}